\documentclass[a4paper,onecolumn,11pt]{quantumarticle}
\pdfoutput=1
\usepackage[utf8]{inputenc}
\usepackage[english]{babel}
\usepackage[T1]{fontenc}
\usepackage{amsmath}
\usepackage{hyperref}
\usepackage{breqn}

\usepackage{tikz}
\usepackage{lipsum}

\begin{document}

\title{Schrieffer-Wolff Transformation on IBM Quantum Computer}

\author{Rukhsan Ul Haq}
\affiliation{IBM Quantum,IBM Bangalore, India}
\author{ Basit Iqbal}
\affiliation{ Qkrishi, India}
\author{Mohsin Ilahi}
\affiliation{AMU, India}
\author {Baseer Ahmad}
\affiliation{Kashmiri Institute of Mathematics Sciences, India}
\author{Nazama}
\affiliation{Kashmiri Institute of Mathematics Sciences, India}

%\author{Lídia del Rio}
%\affiliation{Institute for Theoretical Physics, ETH Zurich, 8093 Zurich, Switzerland}
%\orcid{0000-0002-2445-2701}
%\author{Christian Gogolin}
%\email{latex@quantum-journal.org}
%\homepage{http://quantum-journal.org}
%\orcid{0000-0003-0290-4698}
%\thanks{You can use the \texttt{\textbackslash{}email}, \texttt{\textbackslash{}homepage}, and \texttt{\textbackslash{}thanks} commands to add additional information for the preceding \texttt{\textbackslash{}author}. If applicable, this can also be used to indicate that a work has previously been published in conference proceedings.}
%\affiliation{Covestro Deutschland AG, Kaiser-Wilhelm-Allee 60, 51373 Leverkusen, Germany}
%\author{Marcus Huber}
%\affiliation{Institute for Quantum Optics \& Quantum Information (IQOQI), Austrian Academy of Sciences, Boltzmanngasse 3, Vienna A-1090, Austria}
%\orcid{0000-0003-1985-4623}
%\author{Christopher Granade}
%\affiliation{Microsoft Research, Quantum Architectures and Computation Group, Redmond, WA 98052, USA}
%\author{Johannes Jakob Meyer}
%\affiliation{Dahlem Center for Complex Quantum Systems, Freie Universität Berlin, 14195 Berlin, Germany}
%\orcid{0000-0003-1533-8015}
%\author{Victor V. Albert}
%\affiliation{Institute for Quantum Information and Matter \& Walter Burke Institute for Theoretical Physics, Caltech, Pasadena, CA 91125, USA}
%\orcid{0000-0002-0335-9508}
\maketitle

\begin{abstract}
   Schrieffer-Wolff transformation (SWT) has been extensively used  in quantum many-body physics to calculate the low energy effective Hamiltonian. It provides a perturbative
method to comprehend the renormalization effects of strong correlations in the quantum many-body models. The generator  for Schrieffer-Wolff transformation is
calculated usually by heuristic  methods. Recently, a systematic
and elegant method for the calculation of this extremely significant transformation has been reported~\cite{Haq_1}. Given the huge significance of SWT for many areas including quantum condensed matter physics, quantum optics and  qantum cavity electrodynamics, it is imperative to develop quantum algorithm for carrying out SWT on quantum computer. In this paper, we put forward this quantum algorithm and demonstrate it for single impurity Anderson model (SIAM), thereby arriving at Kondo model as effective Hamiltonian. We implement our quantum algorithm in QisKit and carry out SWT for SIAM on IBM Quantum computers. To the best of our knowledge, this work is the first of its kind to obtain Kondo model from Anderson impurity model using a quantum algorithm. 
\end{abstract}

\section{Introduction } 

Comprehending and controlling quantum many-body
systems is of immense importance in modern theoretical, computational and experimental physics~\cite{Vidal_1,Islam_1,Carleo_1}.
As the Hilbert space grows exponentially, it is challenging to do exact analytical calculations with the increase in
system size. In majority of applications, details about the low-energy features are enough, hence the
description of the low-energy effective Hamiltonian ($H_{eff}$) has
got a significant role in many-body physics~\cite{Soliverez_1}. 

Schrieffer-Wolff transformation(SWT) was initially proposed in ~\cite{Schrieffer_1} to connect Anderson impurity
model with Kondo model in the strong coupling regime. Since then this unitary transformation has been employed excessively
in condensed matter physics. In ~\cite{Bravyi_1} recent review on SW
transformation has been reported. There a mathematically rigorous description of the
transformation has been presented. SWT can be regarded as a degenerate perturbation
theory.For example, due to SWT we can deduce that the Heisenberg model in the strong coupling regime is effectively equivalent to the Fermi-Hubbard model~\cite{MacDonald_1,Fazekas_1}, where it becomes difficult to use approaches from perturbation theory~\cite{ Cleveland_1}. Having being used and developed for myriad quantum problems~\cite{Paaske_1,Issler_1,Hohenester_1,Erlingsson_1,Uchoa_1,Kessler_1,Heikkila_1,Bukov_1,Bukov_2,Zhang_1,Matlack_1,Yan_1,Wurtz_1,Garbe_1,Murakami_1}, SWT has been christened by various names across different fields: In semiconductor physics known as k.p method~\cite{Winkler_1}, in electron-phonon interaction called as Frohlich transformation~\cite{Frohlich_1} and in relativistic  quantum mechanics named as Foldy-Wouthuysen transformation~\cite{Foldy_1}.

SWT transformation is exploited to calculate the effective Hamiltonians of
the models for strongly correlated electron systems as well as various other quantum many-body
systems. It can be interpreted as single shot renormalization process
that projects out the high energy excitations. This then leads to the low energy effective
Hamiltonians that have the signatures of the projected out high energy excitations manifested as renormalized coupling constants. Although SWT is having
perturbative nature but it offers a technique that provides physical insights and intuitions regarding the quantum many-body systems
in the strong coupling regime . One prototypical illustration
is the derivation of the Kondo model as an effective Hamiltonian of the Anderson impurity
model. Later, it was affirmed that
Kondo model leads to the strong coupling fixed point of Anderson impurity model by using numerical renormalization group calculation. Moreover,one of its important aspects is the unitarity unlike some
projective methods used for calculations of the effective Hamiltonians. SWT
as a unitary transformation cuts the clutter of the off-diagonal terms up to the first order and
thus turns out to be a way to diagonalization. 

Recently, quantum computation has attracted ample attention. It has been shown that quantum
computers are able in tackling problems that are
intractable for classical computers~\cite{Shor_1,Arute_1}.So it naturally becomes highly desirable to find an algorithm for this transformation that can be implemented on a quantum computer. In this paper, we for the first time propose a full quantum algorithm for a quantum computer to realize SWT.

The plan of the paper is as follows. In the next section brief
introduction to SWT and review of relevant different
approaches and interpretations is given. In section 3 a systematic
method for SWT is discussed. In section 4, the quantum algorithm for SWT on quantum computer is presented. In Section 5, a test of the quantum algorithm for SWT on a  discrete version of two-site single  impurity Anderson
model(SIAM) is done analytically. This will be followed by a test on a general version of SIAM analytically in Section 6. After that these two models will be simulated on a quantum computer provided by IBM Qiskit and the results will be shown in Section 7. Finally we shall summarise and conclude in Section 8.

\section{Schrieffer-Wolff Transformation}
In this section we shall briefly introduce SWT and review some progress in the direction of its generalization.
Unitary transformations are the typical methods for diagonalization in quantum
mechanics as well as condensed matter physics~\cite{Wagner_1}. In the unitary transformation one
goes to the basis where the given Hamiltonian turns out to be diagonal. This 
leads to diagonalization as single step process. However, such process is impossible for every
Hamiltonian. Thus in the foregoing case one attempts diagonalization of the Hamiltonian by a perturbative
way. It is still possible to employ unitary transformation to do diagonalization of difficult Hamiltonians. SWT is one very important method to tame such Hamiltonians. It not only does diagonalization of the
Hamiltonian by a perturbative way in that case it behaves as a unitary perturbation theory
but it also does renormalization of the the parameters present in the Hamiltonian and thus can manifest itself as a type of renormalization method. SWT in the foregoing sense is employed
to obtain an effective Hamiltonian for the given Hamiltonian and thus leads us to a
special region of a Hamiltonain within its parameter space. It is precisely the way  SWT
 was first applied in~\cite{Schrieffer_1}.
SWT as discussed above can be used as a unitary transformation. Thus finding an appropriate 
unitary operator is vital that could either completely diagonalize the Hamiltonian or to a desired
 order of perturbation.

\begin{equation}
H'=U^{\dagger}HU ~~~~~~~~~~~H'=e^{S}He^{-S}.
\end{equation}
Here S represents the generator of the transformation. It is an anti-Hermitian operator.
Normally one desires of this unitary transformation to nullify the off-diagonal terms upto the first
order by satisfying the condition as given below.

\begin{equation}
[S,H_0]=H_v.
\end{equation}
On elaborating the operator exponential by BCH formula one obtains series expansion pertaining to
the transformed Hamiltonian $H^{`}$ as follows
\begin{equation}
H'=H_0+\frac{1}{2}[S,H_v]+\frac{1}{3}[S,[S,H_v]]+...
\end{equation}
Here $H_0$ and $H_v$ are diagonal and off-diagonal pieces of the Hamiltonian $H$ respectively. As the
off-diagonal piece is nullified up-to the first order, thus the effective Hamiltonian up-to the
second order turns out to be
\begin{equation}
H_{eff}=H_0+\frac{1}{2}[S,H_v].
\end{equation}
SWT has become very significant transformation since the day it was first proposed few decades ago. Due to this reason it has been generalized in different ways. One remarkable generalization has been done by Wegner~\cite{Wegner_1}
as well as Glazek and Wilson~\cite{Glazek_1} independently. The new procedure has been named by Wegner as Flow
Equation Method whereas it has been labelled as Similarity Renormalization by Glazek and Wilson. The unitary transformation is  again employed in Flow Equation Method. However, it is performed in
a continous way as the generator relies upon the flow parameter. The connection of SWT with Flow Equation Method has been illuminated in \cite{Kehrein_1}. The generalization of SWT has been extended to dissipative quantum systems recently \cite{Kessler_1}.

Moreover, the procedure of SWT that has been discussed so far 
uses a generator to accomplish the transformation. There exists a method for carrying out the transformation where one employs Projection operators~\cite{Hewson_1} or Hubbard operators~\cite{Fazekas_1}. Utilizing projection operator method one gets the effective Hamiltonian. However, it fails to reproduce
every term that one obtains in SWT or generator method. Projection operator can be considered in line with SWT due to its  renormalization character where as 
Hubbard operator method is specifically applicable to Hubbard model.
As is evident in the above discussion for evaluating $H_{eff}$ one needs to find a generator that is very crucial and lies at the heart of SWT or generator method. It is natural to have a systematic method for finding such an SWT generator. This is precisely the subject matter of the next section.

\section{Systematic Method for Schrieffer-Wolff Transformation}
In this section we shall try to briefly review the progress in the development of systematic methods for doing Schrieffer-Wolff Transformation both analytically and quantum computationally.

In \cite{Haq_1}, a new systematic method for finding generator used in doing SWT has been reported. This analytic method is simple yet powerful. The method leads to the required generator for SWT by just following the two steps as mentioned below
\begin{itemize}
    \item In the first step commutator $[H_{o}, H_{v} ]$ is evaluated and named as $\eta$, where $ H_{o}$ and $H_{v} $ denote the diagonal and  off-diagonal part of a given Hamiltonian $H$ respectively.
    \item In the second step condition is applied  on $\eta$ for removing its off-diagonal part upto first order. In order to do that coefficients are kept undetermined and are then determined by above condition.  $\eta$ is supposed to satisfy $ [\eta, H_{o}] = - H_{v}$ to be the generator of the transformation. The preceding constraint identifies the coefficients and one gets the generator for SWT of the desired Hamiltonian.
\end{itemize}
This  systematic method has been used to evaluate the generator of SWT for the classic model SIAM, in which SWT was employed for the first time~\cite{Schrieffer_1}. Moreover, the calculation of generator for doing SWT to other models like Periodic Anderson Model, Anderson Holstein Model, Frohlich Hamiltonian and Jaynes-Cummings Model has been done analytically showing the utility and versatility of this novel and elegant method~\cite{Haq_1}.
A question naturally arises. If we have now a systematic method that can find generator for SWT analytically, should not we have algorithms that can do SWT on quantum computer to make life even easier. As the fault-tolerant universal computer~\cite{Shor_2,Gottesman_1} might not be possible in immediate future, noisy intermediate-scale quantum (NISQ) hardwares are feasible options for applications in different fields that include many-body quantum physics~\cite{Preskill_1}. Beginning with the Variational Quantum Eigensolver (VQE)~\cite{Peruzzo_1}, that's a Variational Quantum Algorithm (VQE)~\cite{Cerezo_1} to evaluate by employing shallow circuits the ground state of a desired Hamiltonian, various beneficial quantum algorithms for NISQ hardware have been put forward~\cite{Kandala_1,Li_1,Farhi_1,Xu_1}. Very recently~\cite{Zhang_2} have reported some more progress in this direction. In that paper, the authors have proposed a quantum algorithm for realising SWT. They have also proposed a hybrid quantum-classical  algorithm  for determining the effective Hamiltonian on a hardware available in NISQ era and have applied the algorithm for Heisenberg model.

However, we don't yet have a full fledged general quantum algorithm for SWT that can be directly implemented on quantum computer irrespective of the model under consideration. To fill this gap, we have developed one such  quantum  algorithm that can be directly implemented on quantum hardware. This quantum algorithm is inspired by the systematic method~\cite{Haq_1} and is discussed in the next section.

\section{  Quantum Algorithm for SWT }
In this section we shall spell out SWT as a quantum algorithm for a general many body Hamiltonian that can be directly implemented on a quantum computer. The quantum algorithm for SWT is given in the following steps.

1. First step is to choose a desired many-body Hamiltonian $H$ for which SWT has to be simulated on a quantum computer.

2. Then the fermionic operators in the chosen Hamiltonian have to be written in a language that quantum computer understands. This language is qubit language. That means in other words writing fermionic operators as unitary operators that can be used as quantum gates on a quantum computer. For that one has to map the fermionic operators to qubit operators like for example Pauli strings. Many such mappings are available like Jordan-Wigner(JW) Mapping, Brayvi-Kitaev Mapping and Parity mapping to name a few.

3. Once we have converted the chosen Hamiltonian in qubit language by a particular mapping, we can write qubit Hamiltonian mathematically  as follows
\begin{equation}\label{qubit_hamiltonian}
    H = H_q
\end{equation}

where $H_q$ is the chosen Hamiltonian in the qubit language.

4. Now the qubit Hamiltonian obtained in Eq.(\ref{qubit_hamiltonian}) can be dissected into two parts, viz., diagonal part and an off-diagonal part as follows 
\begin{equation}
  H_q= H_{q}^{o} + H_{q}^{v} 
\end{equation}
where,
$ H_{q}^{o}$ and $H_{q}^{v} $ denote the diagonal and  off-diagonal part of qubit Hamiltonian $H_q$ respectively. All terms of diagonal part and an off-diagonal part commute within themselves but not across.

5. After that one can proceed for an evaluation of the ansatz as proposed in~\cite{Haq_1}. This ansatz makes life easier by finding the generator for SWT systematically in the qubit language and is determined by the following commutator
\begin{equation}
  {\eta}_q=[H_{q}^{o}, H_{q}^{v} ].
\end{equation}
At this stage ${\eta}_q$ gives the form of $S_q$, that is, qubit generator for quantum SWT with unknown coefficients.

6. To get the final form of $S_q$ obtained in previous step with known coefficients, it needs to be processed further so that coefficients are determined. For that an evaluation of the following commutator is required
\begin{equation}
    [S_q, H_{q}^{o}] = - H_{q}^{v}.
\end{equation}
The above constraint helps one to obtain the required qubit generator with known coefficients needed for quantum simulation of SWT.

7. Finally we employ the quantum generator $S_{q}$ evaluated at the end of step 6 in the second term of the effective qubit Hamiltonian as follows
\begin{equation}\label{H_q_eff}
  H_{q}^{eff}= H_{q}^{o} +\frac{1}{2}[S_{q}, H_{q}^{v}].
\end{equation}
Where
$H_{q}^{eff}$ is the required effective qubit Hamiltonian upto second order.

In the following sections our concern will be to implement this algorithm. First we shall describe it analytically followed by implementation on quantum simulator. To demonstrate all this we need to take a prototypical model. One such model is the SIAM. So we discuss it analytically in the next section by taking first a simpler version of it.
\section{Analytical Calculations to Test Quantum SWT Algorithm}
It would be appropriate at this point to demonstrate analytically the quantum SWT algorithm proposed in previous section on a simple model. The first step would be then to choose one such model. In this section we shall do analytical calculations for quantum SWT generator and calculate qubit effective Hamiltonian upto second order $H_{q}^{eff}$  for a simple version of SIAM that consists of one electron in conduction band and one in impurity level. The Hamiltonian for this 2-site model looks like the following

\begin{equation}
  H=U n_{1\downarrow}n_{1\uparrow} - \mu \sum_{\sigma} n_{1\sigma} + \sum_{\sigma} \epsilon_{2}c_{2\sigma}^{\dagger}c_{2\sigma} + \sum_{\sigma} V (  c_{1\sigma}^{\dagger}c_{2\sigma} + c_{2\sigma}^{\dagger}c_{1\sigma} ).
\end{equation}
Now according to Step 2 of the quantum SWT algorithm, we need to map this Hamiltonian to qubit language by using any of the qubit mappings available. Here we use JW mapping and the Hamiltonian in qubit language turns out to be the following.

\begin{equation}
    H_q=\frac{U}{4}\sigma_{1}^{z}\sigma_{3}^{z}+ (\frac{\mu}{2} - \frac{U}{4})(\sigma_{1}^{z}+\sigma_{3}^{z})-\frac{\epsilon_2}{2}(\sigma_{2}^{z}+\sigma_{4}^{z})+\frac{V}{2}(\sigma_{1}^{x}\sigma_{2}^{x} + \sigma_{1}^{y}\sigma_{2}^{y} +\sigma_{3}^{x}\sigma_{4}^{x} +\sigma_{3}^{y}\sigma_{4}^{y}).
\end{equation}
So now onwards as per the quantum algorithm for SWT as spelled out in the last section we proceed as follows. First we split the above qubit Hamiltonian into diagonal and off-diagonal parts as given below
\begin{equation}
      H_q= H_{q}^{o} + H_{q}^{v} 
\end{equation}
where 
\begin{equation}\label{my_first_eqn}
H_{q}^{o} = \frac{U}{4}\sigma_{1}^{z}\sigma_{3}^{z}+ (\frac{\mu}{2} - \frac{U}{4})(\sigma_{1}^{z}+\sigma_{3}^{z})-\frac{\epsilon_2}{2}(\sigma_{2}^{z}+\sigma_{4}^{z})
\end{equation}
and \begin{equation}\label{my_second_eqn}
    H_{q}^{v}= \frac{V}{2}(\sigma_{1}^{x}\sigma_{2}^{x} + \sigma_{1}^{y}\sigma_{2}^{y} +\sigma_{3}^{x}\sigma_{4}^{x} +\sigma_{3}^{y}\sigma_{4}^{y}).
\end{equation}
Next step is to calculate $\eta_q$. For that we need to evaluate the following commutator
\begin{equation}
  {\eta}_q=[H_{q}^{o}, H_{q}^{v} ].
\end{equation}
Using Eq.(\ref{my_first_eqn}) and Eq.(\ref{my_second_eqn}), the commutator turns out to be as follows
\begin{equation}
  {\eta}_q=\frac{i UV}{4}(\sigma_{3}^{z}(\sigma_{1}^{y}\sigma_{2}^{x} - \sigma_{1}^{x}\sigma_{2}^{y})+\sigma_{1}^{z}(\sigma_{3}^{y}\sigma_{4}^{x} - \sigma_{3}^{x}\sigma_{4}^{y})) +(i V(\frac{\mu}{2}-\frac{U}{4})+ \frac{i \epsilon_2 V}{2} )(\sigma_{1}^{y}\sigma_{2}^{x} - \sigma_{1}^{x}\sigma_{2}^{y}+\sigma_{3}^{y}\sigma_{4}^{x} - \sigma_{3}^{x}\sigma_{4}^{y}).
\end{equation}
According to the quantum SWT algorithm, the above equation suggests the form of the qubit generator $S_q$ for SWT as follows
\begin{equation}\label{my_third_eqn}
  S_q=A(\sigma_{1}^{y}\sigma_{2}^{x} - \sigma_{1}^{x}\sigma_{2}^{y})(1-\sigma_{3}^{z}) +B(\sigma_{3}^{y}\sigma_{4}^{x} - \sigma_{3}^{x}\sigma_{4}^{y})(1-\sigma_{1}^{z}).
\end{equation}
To determine the coefficients A and B, the constraint as given in Step 6 of the quantum algorithm has to be applied as follows
\begin{equation}
    [S_q, H_{q}^{o}] = - H_{q}^{v}.
\end{equation}
Using Eq.(\ref{my_first_eqn}), Eq.(\ref{my_second_eqn}) and Eq.(\ref{my_third_eqn}) in above constrain and then after evaluation, we get
\begin{equation}
    A=\frac{-iV}{4U(1-\sigma_{3}^{z})},
\end{equation}
\begin{equation}
    B=\frac{-iV}{4U(1-\sigma_{1}^{z})}.
\end{equation}
Once we know $S_q$ completely, the last step of calculating $H_{q}^{eff}$ up to second order is straightforward. We only need to plug in the following commutator result in the second term of $H_{q}^{eff}$ of Eqn.(\ref{H_q_eff})
\begin{equation}
  [S_{q}, H_{q}^{v}]=\frac{V^2}{4U}(2(\sigma_{1}^{z}+\sigma_{3}^{z})-2(\sigma_{2}^{z}+\sigma_{4}^{z})).
\end{equation}

This completes analytically the demonstration of quantum algorithm for SWT on the simple version of 2-site SIAM. It can be extended to multiple electrons in the conduction band and forms the subject matter of the next section.

\section{ Quantum Algorithm for Anderson Impurity Model}
In this section we shall do analytical calculations for quantum SWT generator and calculate qubit effective Hamiltonian upto second order $H_{q}^{eff}$  for a multiple site SIAM in 1 dimension. The Hamiltonian for this $N$-site model looks like the following

\begin{equation}
  H=\sum_{i=1,\sigma}^{N} V_{i} (c_{0\sigma}^{\dagger}c_{i\sigma} + c_{i\sigma}^{\dagger}c_{0\sigma} )+U n_{0\downarrow}n_{0\uparrow} + \sum_{i=0,\sigma}^{N}(\epsilon_{i}-\mu)n_{i\sigma}.
\end{equation}
The JW mapping allows us to transform the
fermionic Hamiltonian into the following qubit Hamiltonian language

\begin{dmath}
  H_q=\sum_{i=1}^{N} \frac{V_{i}}{2} (X_0Z_1...Z_{i-1}X_i +Y_0Z_1...Z_{i-1}Y_i + X_{N+1}Z_{N+2}...Z_{N+i}X_{N+1+i} + Y_{N+1}Z_{N+2}...Z_{N+i}Y_{N+1+i})\\
+\frac{U}{4}( Z_{0}Z_{N+1}-Z_{0}-Z_{N+1}) + \sum_{i=0}^{N}\frac{(\epsilon_{i}-\mu)}{2}(Z_i+Z_{N+1+i}).
\end{dmath}
Now we shall split the above qubit Hamiltonian into diagonal and off-diagonal parts as given below

\begin{dmath}\label{my_fourth_eqn}
  H_{q}^0=\frac{U}{4}( Z_{0}Z_{N+1}-Z_{0}-Z_{N+1}) + \sum_{i=0}^{N}\frac{(\epsilon_{i}-\mu)}{2}(Z_i+Z_{N+1+i}),
\end{dmath}

\begin{dmath}\label{my_fifth_eqn}
  H_q^{v}=\sum_{i=1}^{N} \frac{V_{i}}{2} (X_0Z_1...Z_{i-1}X_i +Y_0Z_1...Z_{i-1}Y_i + X_{N+1}Z_{N+2}...Z_{N+i}X_{N+1+i} + Y_{N+1}Z_{N+2}...Z_{N+i}Y_{N+1+i}).
\end{dmath}

To calculate $\eta_q$, we need to evaluate the following commutator

\begin{equation}
  {\eta}_q=[H_{q}^{o}, H_{q}^{v} ].
\end{equation}
Using Eq.(\ref{my_fourth_eqn}) and Eq.(\ref{my_fifth_eqn}), the above commutator takes the following form

\begin{dmath}
 {\eta}_q=2i(\frac{U}{4}+\frac{(\epsilon_{i}-\mu)}{2}+\frac{(\epsilon_{0}-\mu)}{2})\sum_{i=1}^{N} \frac{V_{i}}{2} (Z_1...Z_{i-1}(Y_iX_0(1-Z_{N+1})-X_iY_0(1-Z_{N+1})) + Z_{N+2}...Z_{N+i}(Y_{N+1+i}X_{N+1}(1-Z_0)-X_{N+1+i}Y_{N+1}(1-Z_0))).
\end{dmath}
  The above expression implies the form of the qubit generator $S_q$ for SWT as follows
\begin{dmath}\label{my_sixth_eqn}
 S_q=\sum_{i=1}^{N} (AZ_1...Z_{i-1}(Y_iX_0(1-Z_{N+1})-X_iY_0(1-Z_{N+1})) + BZ_{N+2}...Z_{N+i}(Y_{N+1+i}X_{N+1}(1-Z_0)-X_{N+1+i}Y_{N+1}(1-Z_0))).
\end{dmath}
To find the coefficients A and B, the following SWT condition has to be followed  
\begin{equation}
    [S_q, H_{q}^{o}] = - H_{q}^{v}.
\end{equation}
Employing Eq.(\ref{my_fourth_eqn}), Eq.(\ref{my_fifth_eqn}) and Eq.(\ref{my_sixth_eqn}) in above condition gives 
\begin{equation}
    A=\frac{-iV}{4U(1-Z_{N+1})},
\end{equation}
\begin{equation}
    B=\frac{-iV}{4U(1-Z_0)},
\end{equation}
and thus our required generator takes the following form
\begin{dmath}
 S_q=\sum_{i=1}^{N}\frac{-iV}{4U} (Z_i...Z_{i-1}(Y_iX_0-X_iY_0) + Z_{N+2}...Z_{N+i}(Y_{N+1+i}X_{N+1}-X_{N+1+i}Y_{N+1})).
\end{dmath}
After finding $S_q$ the final step of evaluating $H_{q}^{eff}$ up to second order is given by using following commutator result in the second term of $H_{q}^{eff}$ of Eqn.(\ref{H_q_eff})
\begin{dmath}
 [S_q, H_{q}^{v}]=\sum_{i=1}^{N}\frac{V^2}{4U} (Z_i...Z_{i-1}(-2Z_i-2Z_0)).
\end{dmath}

This completes the analytical demonstration of quantum algorithm for SWT of multiple electrons in the conduction band of SIAM. Verifying this quantum algorithm for SWT experimentally for various cases is the next important stage and is discussed in the following section.

\section{Experiments on IBM Quantum Devices}
We implemented our quantum algorithm for SWT in Qiskit and did the experiments on IBM Quantum computers. In this section, we give the details of the our experiments and the results. The Qiskit code snippets for various steps of the quantum algorithm have also been included. These figures are for the case of 2-site(4 qubit) SIAM. The results for 8 qubit SIAM are given in the Appendix A.
%We tested our quantum algorithm for 2-site SIAM which has one conduction electron and one impurity electron and hence gets mapped to 4 qubit system. We also did the test for 4-site SIAM which has three conduction electrons and one imputity electron and hence takes 8 qubits. More conduction electrons can be incorporated if the larger number of qubits are available on the quantum computer. All the various step with QisKit code are shown in below figures. 

%The first step of the quantum algorithm is the fermion-to-qubit mapping which is Jordan-Wigner mapping in our case. Once the SIAM has been mapped to qubit operators, we need to calculate $\eta$ from the commutator of diagonal and off-diagonal terms of the Hamiltonian. Then the generator is being calculated and then used to carry out SWT for the SIAM. All the various step with QisKit code are shown in below figures. These figures are for the case of 2-site(4 qubit) SIAM. The results for 8 qubit SIAM are given in the Appendix A. 

%In this section we present the results of some of the representative cases when we implement proposed quantum algorithm of SWT on quantum simulator for SIAM. First four figures are for 4 qubit SIAM that correspond to various important steps of quantum SWT algorithm when implemented on a quantum simulator. Similarly, remaining figures represent the most important steps for 8 qubit case.

\begin{figure}[h!]
\centering
\includegraphics[width=1.0\textwidth]{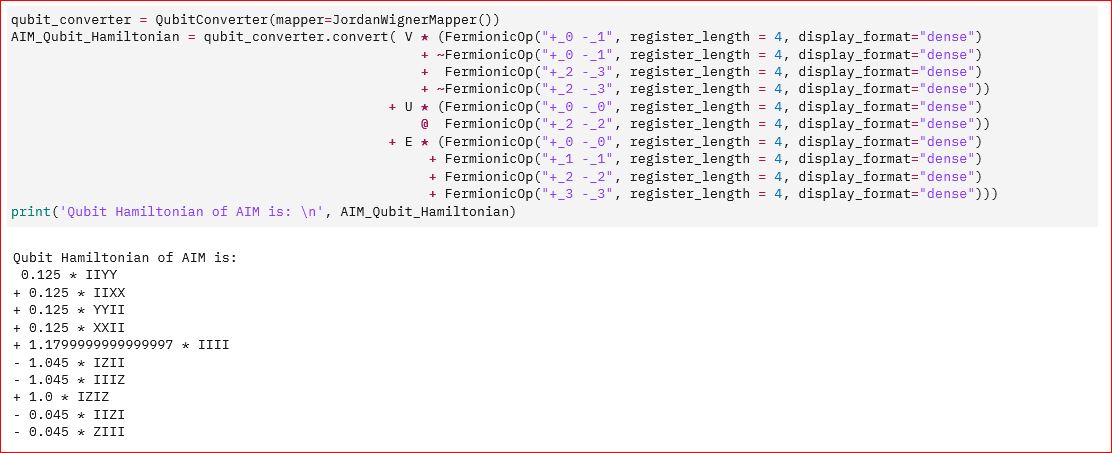}
\caption{\label{fig:H_4q_JWT}4 qubit Hamiltonian after Jordan-Wigner Mapping of 2-site SIAM Hamiltonian.}
\end{figure}

\begin{figure}[h!]
\centering
\includegraphics[width=1.0\textwidth]{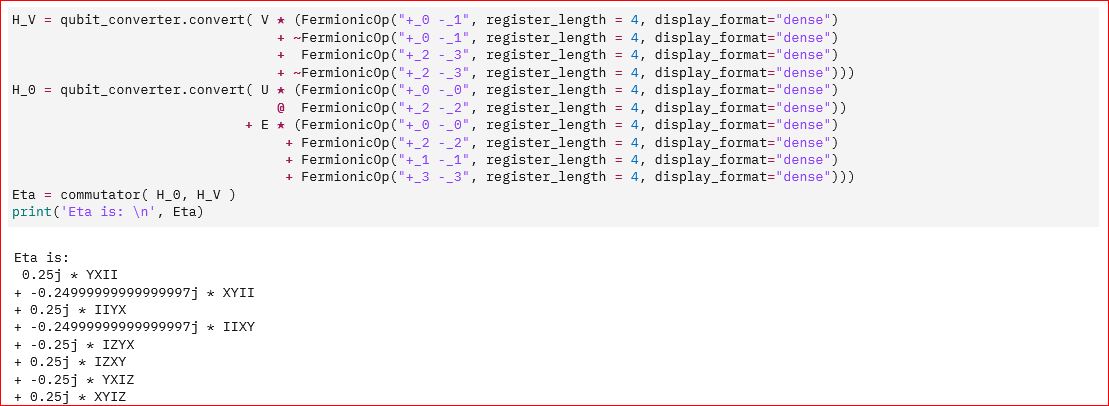}
\caption{$\eta_q=[H_{q}^{0},H_{q}^{v}]$ for 4 qubit Hamiltonian of 2-site SIAM 2-site SIAM Hamiltonian.}
\end{figure}

\begin{figure}[h!]
\includegraphics[width=1.0\textwidth]{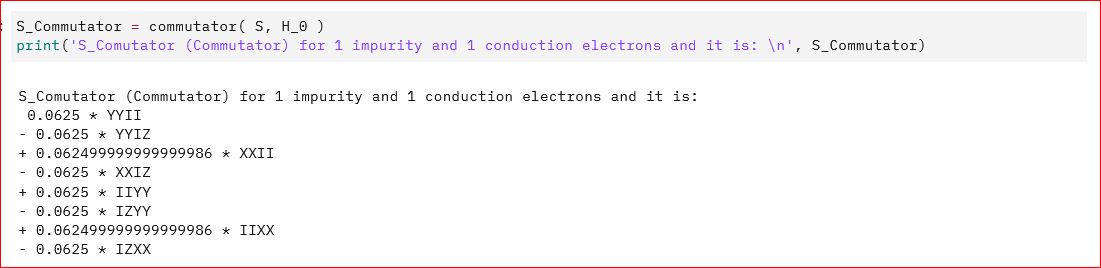} 
\caption{Commutator $[S_q,H_{q}^{0}]$ for 4 qubit Hamiltonian of 2-site SIAM Hamiltonian.}
\end{figure}

\begin{figure}[h!]
\includegraphics[width=1.0\textwidth]{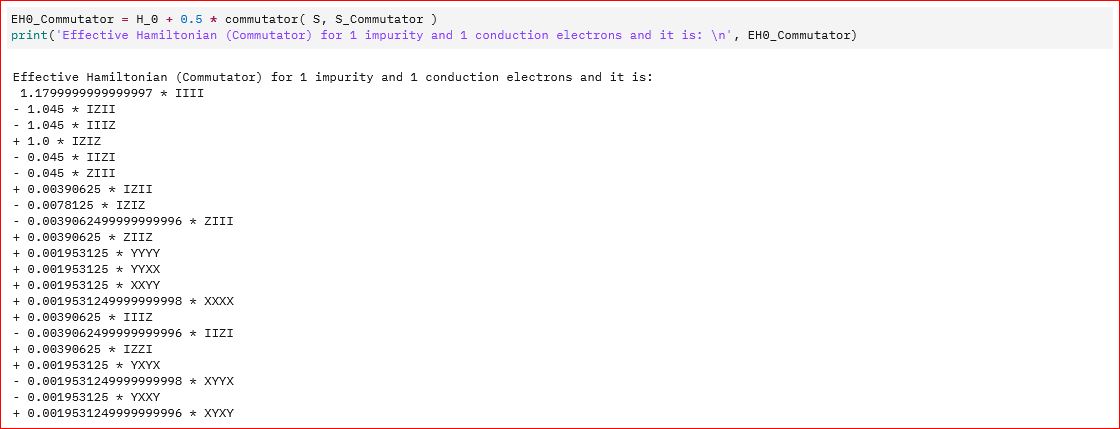} 
\caption{4 qubit Effective Hamiltonian $H_{q}^{eff}$ for 2-site SIAM Hamiltonian after SWT.}
\end{figure}

\section{  Conclusion and Discussion }
Schrieffer-Wolff transformation is extensively used in diverse areas of quantum many-body physics including quantum condensed matter physics, quantum optics, quantum electrodynamics and has found applications in quantum complexity theory. This transformation is routinely used by researchers to calculate the low energy effective Hamiltonians corresponding to interesting fixed points which arise due to the effect of strong correlations in these systems. Therefore, it becomes natural to have quantum algorithm for carrying out Schrieffer-Wolff transformation on quantum computer. In this work, we have demonstrated such an algorithm and have reproduced the historical mapping of Anderson impurity model to Kondo model which was shown in the very first application of SWT. Our quantum algorithm is not restricted to Anderson impurity model, rather it can applied to a broad class of quantum many-body Hamiltonians which share similar structure to Anderson Impurity model. We hope our work will open up new directions for carrying out SWT for various quantum many-body models using quantum computers.

\bibliographystyle{plain}

\onecolumn\newpage
\appendix

\section{First section of the appendix}
\begin{figure}[h!]
\includegraphics[scale=0.6]{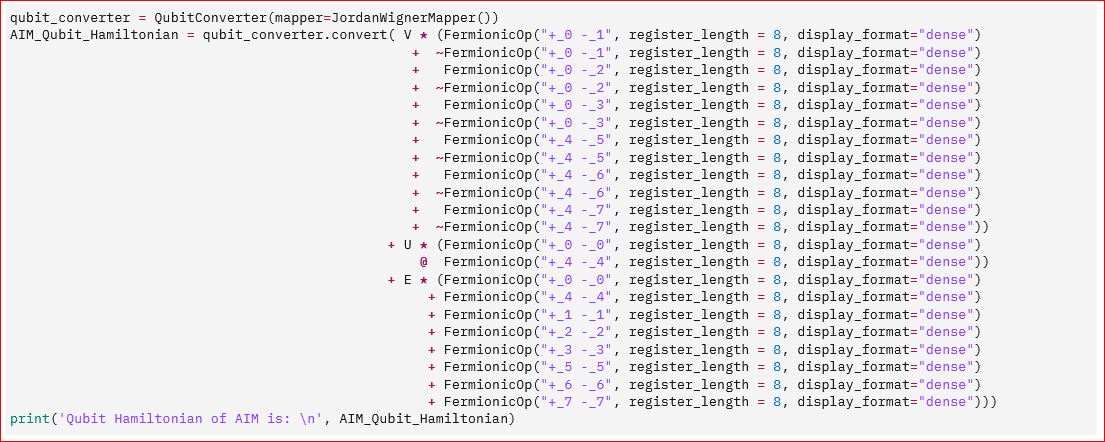} 
\caption{Jordan-Wigner Mapping code for 8 qubit Hamiltonian of 4-site SIAM.}
\end{figure}

\begin{figure}[h!]
%\centering
\includegraphics[scale=1]{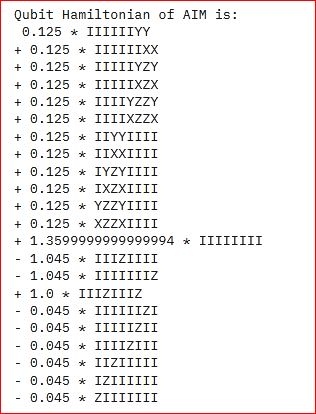}
\caption{\label{fig:H_8q_JWT_2} 8 qubit Hamiltonian after Jordan-Wigner Mapping of 4-site SIAM Hamiltonian.}
\end{figure}

\begin{figure}[h!]
\includegraphics[scale=0.6]{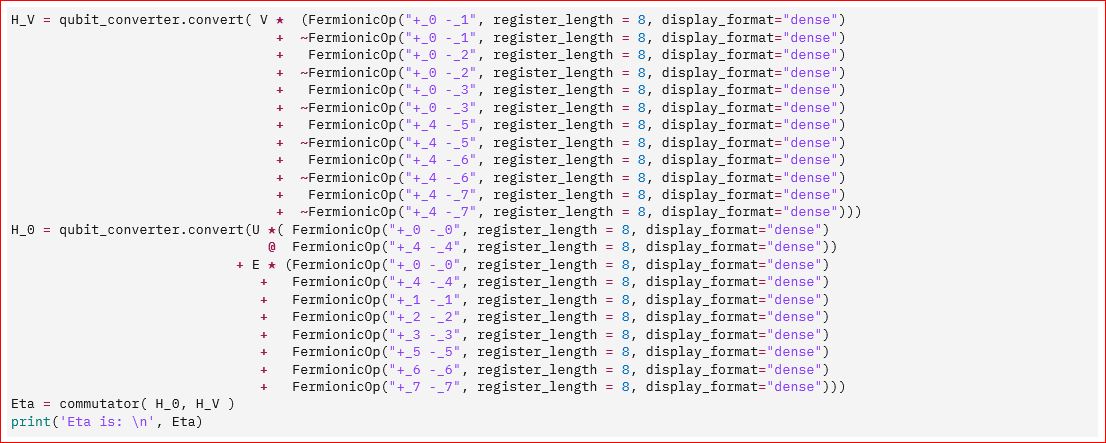}
\caption{\label{fig:Eta_8q_1}Code for $\eta_q=[H_{q}^{0},H_{q}^{v}]$ for 8 qubit Hamiltonian of 4-site SIAM Hamiltonian.}
\end{figure}

\begin{figure}[h!]
%\centering
\includegraphics[scale=1]{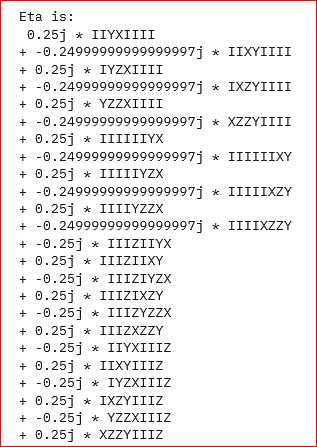}
\caption{\label{fig:Eta_8q_2}$\eta_q=[H_{q}^{0},H_{q}^{v}]$ for 8 qubit Hamiltonian of 4-site SIAM Hamiltonian.}
\end{figure}

\begin{figure}[h!]
\includegraphics[scale=0.6]{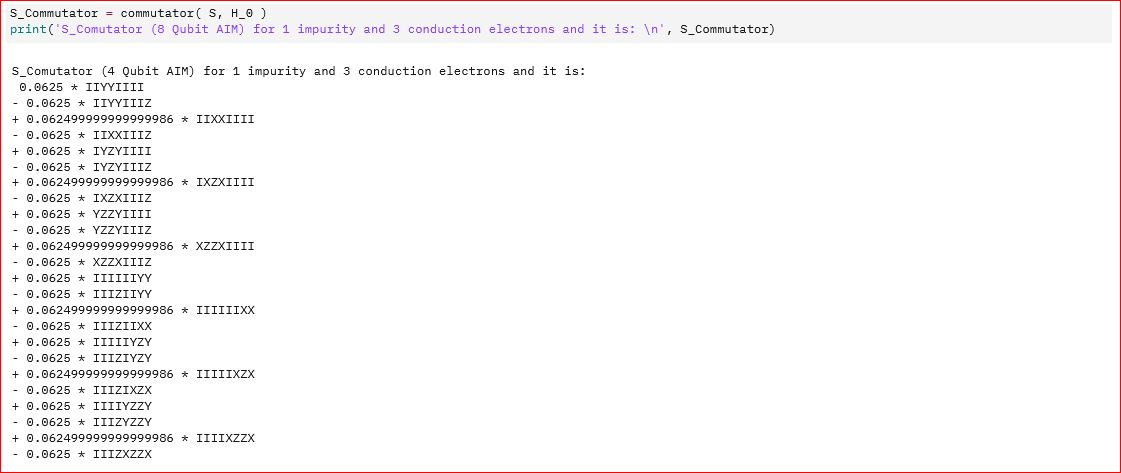}
\caption{\label{fig:S_H_0_8q}Commutator $[S_q,H_{q}^{0}]$ for 8 qubit Hamiltonian of 4-site SIAM Hamiltonian.}
\end{figure}

\begin{figure}[h!]
\includegraphics[scale=0.6]{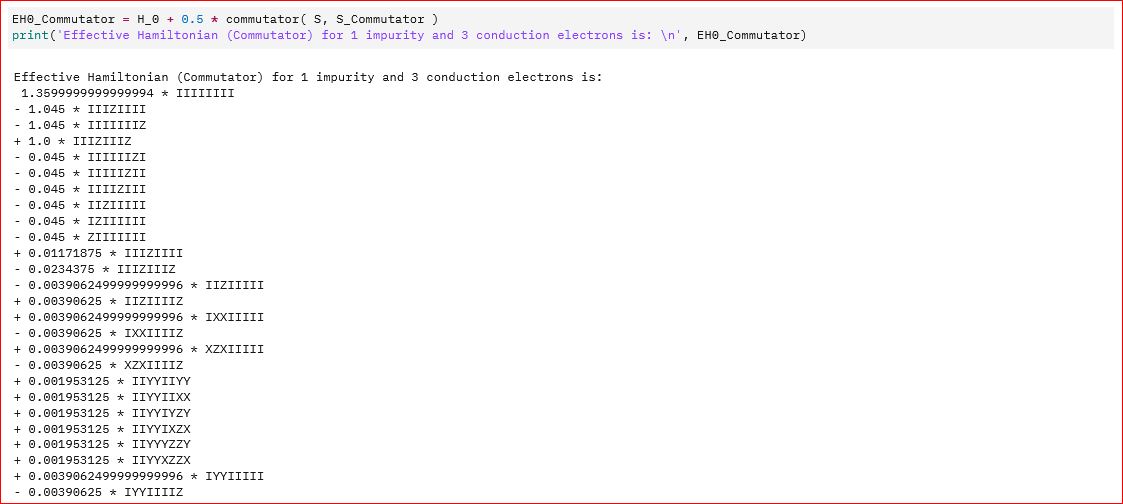}
\caption{\label{fig:H_eff_8q_1}8 qubit Effective Hamiltonian $H_{q}^{eff}$ for 4-site SIAM Hamiltonian after SWT.}
\end{figure}

\begin{figure}
%\centering
\includegraphics[scale=1]{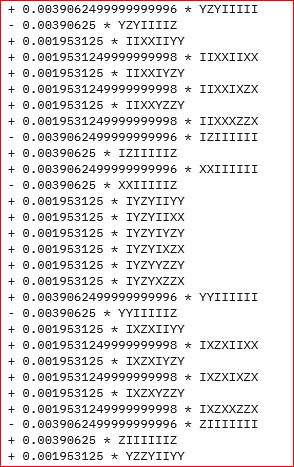}
\caption{\label{fig:H_eff_8q_2}Remaining part of 8 qubit Effective Hamiltonian $H_{q}^{eff}$ for 4-site SIAM Hamiltonian after SWT.}
\end{figure}

\begin{figure}[h!]
%\centering
\includegraphics[scale=1]{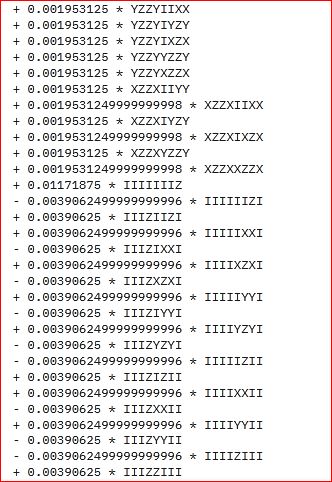}
\caption{\label{fig:H_eff_8q_3}Remaining part of 8 qubit Effective Hamiltonian $H_{q}^{eff}$ for 4-site SIAM Hamiltonian after SWT.}
\end{figure}

\begin{figure}
%\centering
\includegraphics[scale=1]{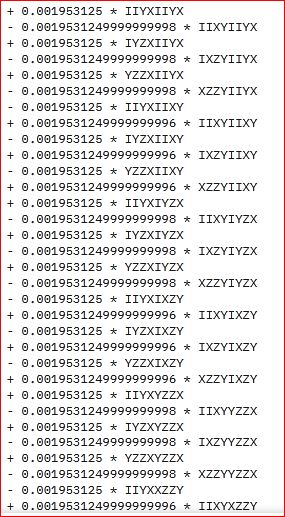}
\caption{\label{fig:H_eff_8q_4}Remaining part of 8 qubit Effective Hamiltonian $H_{q}^{eff}$ for 4-site SIAM Hamiltonian after SWT.}
\end{figure}

\begin{figure}
%\centering
\includegraphics[scale=1]{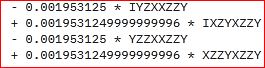}
\caption{\label{fig:H_eff_8q_5}Remaining part of 8 qubit Effective Hamiltonian $H_{q}^{eff}$ for 4-site SIAM Hamiltonian after SWT.}
\end{figure}

\end{document}